\documentclass[sn-mathphys-ay]{sn-jnl}

\usepackage{graphicx}%
\usepackage{multirow}%
\usepackage{amsmath,amssymb,amsfonts}%
\usepackage{amsthm}%
\usepackage{mathrsfs}%
\usepackage[title]{appendix}%
\usepackage{xcolor}%
\usepackage{textcomp}%
\usepackage{manyfoot}%
\usepackage{booktabs}%
\usepackage{algorithm}%
\usepackage{algorithmicx}%
\usepackage{algpseudocode}%
\usepackage{listings}%

\theoremstyle{thmstyleone}%
\theoremstyle{thmstyletwo}%

\theoremstyle{thmstylethree}%

\newcommand{\sm}[1]{{\mbox{{\scriptsize #1}}}}

\newcommand{\be}{\begin{equation}}
\newcommand{\ee}{\end{equation}}
\newcommand{\bea}{\begin{eqnarray}}
\newcommand{\eea}{\end{eqnarray}}
\newcommand{\bdm}{\begin{displaymath}}
\newcommand{\edm}{\end{displaymath}}
\newcommand{\beanonum}{\begin{eqnarray*}}
\newcommand{\eeanonum}{\end{eqnarray*}}

\newcommand{\corr}[1]{#1}

\newcommand{\cm}{\rm{cm}}

\newcommand{\g}{\rm{g}}

\newcommand{\s}{\mbox{s}}

\newcommand{\kb}{k_{\sm{B}}}

\newcommand{\afs}{\alpha}

\newcommand{\vek}[1]{{\bf{#1}}}

\newcommand{\va}{\vek{a}}

\newcommand{\vr}{\vek{r}}

\newcommand{\vF}{\vek{F}}

\begin{document}

\title[Gravity, Fine-Structure Constant and Natural Units]{
  \begin{center}
    Gravity, Fine-Structure Constant\\ and Natural
    Units
  \end{center}
  -- some Thoughts based on Dimensional Analysis --}


\author{\fnm{Robi} \sur{Banerjee}}\email{banerjee@hs.uni-hamburg.de}

\affil{\orgdiv{Hamburger Sternwarte}, \orgname{Universit\"at
    Hamburg}, \orgaddress{\street{Gojenbergsweg 112}, \city{Hamburg}, \postcode{21029}, \country{Germany}}}


\abstract{Here we discuss direct links of the number of fundamental
  dimensions to the fundamental natural constants using simple
  arguments of dimensional analysis \corr{based on Maxwell's
    dimensions length (L), time (T) and mass (M) as well as the
    constants $G$, $c$, $\hbar$ and $e$}. We find that the \corr{form}
  of the fine-structure constant is a direct consequence of this
  connection. Additionally, our approach emphasises that gravity is a
  quite distinct area of physics which is not yet successfully
  quantised, i.e. not yet combined with quantum mechanics.  We also
  discuss different unit systems based on dimensional analysis and
  natural constants.}


\maketitle

\section{Introduction}

Physics, unlike math, which is the appropriate language to phrase
physical laws, is necessarily connected to units to describe the real
world. Physics allows us to determine the phenomena of nature with precession,
such as the motions of pendulums and planets around the sun or the
spectra of light emitted by atoms. To measure these phenomena we need
units to quantify them. In physics, the generalisation of units are
dimensions which are independent of the specific unit system used by
different societies or are defined within institutionalised
agreements, like the International System of Units (SI).  For instance
the dimension of length can be measured in units of meters (SI),
parsec (astronomy) or in Angstroms, which is the typical size of an
atom. All physical phenomena, and hence nature itself, are connected
to such dimensions. Despite the complexity of nature, we need only
{\em three} fundamental dimensions to describe the real
world. \corr{Here, we adopt the three fundamental dimensions $L$,$T$
  and $M$ as a convenient basis for the present discussion.}


An illuminating discussion on these fundamental dimensions of
nature can be already found in Maxwell's famous text book 'A Treatise on
Electricity and Magnetism' (the first modern formulation of
electrodynamics) in the preliminary chapter {\it On the measurement of
  quantities} \citep{Maxwell1873}. There, Maxwell pointed out
that physics only requires {\bf three fundamental dimensions}. Maxwell
identified these as
\begin{itemize}
\item LENGTH ($L$)
\item TIME ($T$)
\item MASS ($M$)
\end{itemize}
Consequently, all known physical laws and quantities can be expressed
in combinations and powers of these fundamental dimensions. Similar to
different unit systems, the specific definition of the fundamental
dimensions is not unique. One could also imagine using a system with
the dimensions of Energy ($E$), Velocity ($V$) and Length ($L$) or Power
($P$), Energy ($E$) and Force ($F$). Thus, fundamental dimensions can
be considered as key information to quantify physical phenomena, which
eventually can be measured with specific units. In this context, the
only invariable property of physics is, that there are only {\em
  three} fundamental dimensions necessary to describe nature.  This
fact is based on the physical laws and the necessity of dimensional
homogeneity of equations. For instance, Newton's equation of motion
has the dimension of $M\,L\,T^{-2}$, Schr\"odinger's equation has the
dimension of Energy, i.e. $M\,L^2\,T^{-2}$, and the same applies to
the first and second laws of thermodynamics.
As a consequence, the {\it number} of the fundamental dimensions can not be
reduced. This is intuitively evident, as for the description of an
event (and hence, a physical process) one needs the information on
'where' ($=L$), 'when' ($=T$) and 'what' ($=M$). 

The fact that three fundamental dimensions are sufficient to describe
nature is \corr{noteworthy given the complexity of the real
  world.}  Sticking with Maxwell's definition of the fundamental
dimensions, $L$, $T$ and $M$, we have a good intuition of spatial
distances quantified by $L$ and also time $T$. For instance, we can
describe the motions of objects or arrange a meeting with those two
dimensions. However, we also know that objects have internal
properties like material composition, phase state (solid, liquid,
gaseous), temperature or charge. Nevertheless, all the internal
properties can be quantified by one more dimension like mass $M$ in
combination with the other two $L$ and $T$. Again, this fact reflects
the simplicity of physics (in terms of a few basic laws of nature) due
to its overall universality and invariability.

\corr{This fact is also reflected in the SI system. Although the SI
  formally consists of seven base units, only three are independent.}
Two SI units, namely 'mol' and 'kelvin' are a defined number and a
specific definition of energy, respectively (see also
Sec.~\ref{sec:phys_consts} on alternative units and the Boltzmann
constant $\kb$), and are not considered to be fundamental units. The
units 'coulomb', 'ampere' and 'candela' can be derived from the units
'meter', 'second' and 'kilogram'. For instance, one ampere is defined
to be a charge flow, i.e. a fixed amount of charge per 'second', where
charge is quantified with the units 'meter', 'second' and 'kilogram'
(see also Eq.~(\ref{eq:e_dims})). The light intensity unit 'candela'
is just a power, which units are again based on 'meter', 'second' and
'kilogram'.

Another unexplained fact of physics, i.e. nature, is the existence of
fundamental natural constants. Indeed, the physical constants, like
the gravitational constant $G$, the speed of light $c$ or Planck's
constant $\hbar$ are constant in space and time and independent on the
environment. At least, so far there is no indication that these might
be different, for instance, shortly after the Big Bang some 14 billion
years ago where the Universe was incredibly dense and hot \citep[see,
e.g.,][on constancy of natural constants]{Chiba2011, Fritzsch2011}.

Unlike for the number of fundamental dimensions, there is no
general agreement on the {\it number} of fundamental physical constants
\citep[see e.g.,][and references herein]{Matsas24}. Nevertheless, we argue
that there are only four fundamental physical constants, namely:
\begin{itemize}
\item $G$ the gravitational constant
\item $c$ the speed of light in vacuum
\item $\hbar$ Planck's constant of quantum mechanics\footnote{We use
    the reduced Planck's constant $\hbar$ throughout, where $h$
    appears it is related by $h = 2\pi\hbar$}
\item $e$ the elementary electric charge, i.e. the charge of the electron
\end{itemize}
In section~\ref{sec:phys_consts} we will discuss this assumption in more
detail.

The discussions on fundamental dimensions and fundamental natural
constants are often implicitly linked, in particular to define a
natural unit system and to argue for more or less numbers of
fundamental dimensions and natural constants~\citep[see
e.g.][]{Duff02, Matsas24}. Natural constants can be used to convert dimensions,
e.g. length in time and vice versa with the speed of light, which is a
natural constant. But in this article we will discuss the connections
of fundamental dimensions and fundamental natural constants explicitly
while treating them as independent concepts. 
We explicitly combine both concepts with the arguments of dimensional
analysis. Although, dimensional analysis has its limitations
\citep[see e.g. discussions from the seminal lecture
of][]{Bridgman22}, it is nevertheless a method that can lead to some
deeper insight of the nature of physics\footnote{\cite['It appears,
  therefore, that {\it dimensional analysis is essentially of the
    nature of an analysis of an analysis}'][]{Bridgman22}}.

The following discussion is based on the assumption that only four
fundamental natural constants and three fundamental dimensions exist
in nature. Then of course, one can argue why shouldn't there be only
three natural constants to match the number of fundamental physical
dimensions. This issue was discussed in a number of articles
\citep[see e.g.,][and references herein]{Wutke23} where the trialogue
by Michael Duff, Lev Okun and Gabriele Veneziano sums up the
controversy of this field very well \citep{Duff02}. There, Okun states
that the number of fundamental dimensions are independent of the
number of fundamental natural constants. And indeed, there is no
obvious reason why those numbers should be equal and in fact -- as far
as we know -- the numbers are not equal. But we would still claim
that, in the light of simplicity of physics, it would be most
``natural'' if there would be as many fundamental dimension as
fundamental natural constants. We will discuss the consequences of
this assumption in the subsequent sections.

\section{Fundamental Dimensions and Natural Constants}
\label{sec:dims_consts}

As mentioned in the introduction we need only three fundamental
dimensions to describe the laws of nature. Here, we will stick to
Maxwell's 'units', i.e. length $L$, time $T$ and mass $M$ to
characterise those dimensions. Then the dimensions of the four
fundamental natural constants are:
\bea
\left[ G \right] & = & M^{-1}\,L^3\,T^{-2}
\label{eq:G_dims}\\
\left[ c \right] & = & L\,T^{-1}
\label{eq:c_dims}\\
\left[ \hbar \right] & = & M\,L^2\,T^{-1}
\label{eq:h_dims}\\
\left[ e \right] & = & M^{1/2}\,L^{3/2}\,T^{-1} \quad,
\label{eq:e_dims}
\eea
where we used the square bracket notation to indicate the
dimensionality of quantities\corr{\footnote{\corr{The dimensions for
      the elementary charge $e$ is derived from the form of Coulomb's
    law: $F_C = e^2/r^2 \to [e] = M^{1/2}\,L^{3/2}\,T^{-1}$. It is
  consistent with the SI, where the unit 'Coulomb' is also a derived
  unit.}}}.

Now, let's assume that for every fundamental dimension ($M$, $L$, $T$)
there is only {\em one} fundamental natural constant. That means that one of
the physical constants, $G$, $c$, $\hbar$ and $e$ can be expressed by
means of the others. Using dimensional analysis will give us the
appropriate relation:
\be
\left[G\right]^{x_G}\,[c]^{x_c}\,[\hbar]^{x_h}\,[e]^{x_e} =
M^0\,L^0\,T^0
\label{eq:const_relation}
\ee
to solve for the exponents $x_G$, $x_c$, $x_h$ and $x_e$
\footnote{i.e. solve the coupled set of linear homogenous equations
  \beanonum
  -x_G + x_h + x_e/2 & = & 0 \\
  3 x_G + x_c + 2 x_h + 3/2 x_e & = & 0 \\
  -2 x_G - x_c - x_h - x_e & = & 0
  \eeanonum}.
The only non-trivial solution of this equation results uniquely to
\be
x_G = 0 \, .
\label{eq:nograv}
\ee
\corr{This implies that the gravitational constant does not enter any
  nontrivial dimensionless combination with $c$, $\hbar$, and
  $e$. Within this framework,} only $\hbar$, $c$ and $e$ can be
related to each other and the general solution of
Eq.~(\ref{eq:const_relation}) is: \bea
\left(\frac{[\hbar]\,[c]}{[e]^2}\right)^{x_h} & = & M^0\,L^0\,T^0
\quad
\mbox{, i.e.} \nonumber \\
\left(\frac{\hbar\,c}{e^2}\right)^{x_h} & = & \mbox{number}
\label{eq:fine_structure}
\eea
with an undetermined exponent $x_h$ and
$\hbar\,c/e^2 \approx 137.036$ is a dimensionless number, namely
the famous {\em fine-structure constant} (the actual fine-structure
constant is $\afs = e^2/\hbar\,c \approx 1/137.036$.). This 
result already justifies our concept to link the number of fundamental
dimensions to the natural constants of physics.

The fact that $G$ doesn't contribute to the conversion of physical
constants shows again that not only the gravitational constant $G$ but
the entire gravitational physics is special and a particular
phenomenon of nature.  One could interpret this feature in a way that
gravity is a distinctively separated domain of physics as $G$ can't be
'derived' from the other constants, whereas $\hbar$, $c$ and $e$ are
linked together by the fine-structure constant, which is a mystery by
itself and will be discussed below.

\corr{Please also note, with a different set of natural constants,
  e.g. including masses of particles, like the mass of the electron
  $m_e$, one can construct easily dimensionless constants including
  $G$ such as $\alpha_G \equiv G\,m_e^2/\hbar c$. In
  Sec.~\ref{sec:phys_consts} we address this point and discuss our
  choice of natural constants used for this work.}

\subsection{Gravity}

Ever since Einstein's General Relativity (GR) we know, that gravity is
actually not a force but a consequence of spacetime properties, i.e. a
geometrical phenomenon. In this context together with the fact that
$G$ can not be used to express other physical constants one could
argue that $G$ is ultimately linked to geometry only and is
independent from other physical interactions. \corr{Maybe this is another
reason why gravity remains difficult to reconcile with quantum
mechanics.} Hence so far, there is no fully satisfying theory of quantum
gravity. 


\corr{Based on the link of gravity and geometry, we claim that the
  physical spacetime can not exist without gravity.  Contrary to
  mathematics, where geometry is independent of gravity, in our real
  physical world gravity is a key component.} The physical world is
necessarily based on dimensions and units. We cannot imagine a world
where humans are just dimensionless numbers. On practical reasons,
that wouldn't work, similarly than living creatures don't work in just
two spatial dimensions. We have seen, that $G$ can not be reduced to a
dimensionless number with other natural constants, unlike the
combination of $e$, $\hbar$ and $c$, and hence it is necessarily
linked to fundamental dimensions. Furthermore, we can not build a
natural unit system without $G$, whereas the other constants can be
mutually replaced to define natural unit systems, as we will see in
Sec.~\ref{sec:unit_system}. Therefore, we argue that the physical
world as we know it can not exist without gravity.

\corr{As a consequence of this argument we conclude that gravity
  generates spacetime.}
The simplest connection of spacetime and gravity is given by
Einstein's field equations. And indeed, its solution from cosmological
conditions lead to the spacetime structure we live in. Peculiar to our
Universe is that its energy density today is so low
($\varrho \sim 10^{-29}\,\g\,\cm^{-3}$) that its spacetime property is
almost that of a flat Minkowski spacetime (even the relative deviation
from a flat spacetime at the surface of the Earth is only
$\sim 10^{-9}$, although we experience gravity in our daily
life). Otherwise, the perfectly flat and static Minkowski spacetime is
the only solution of the field equations which does not involve
gravity directly, i.e. $G$ does not appear in the solution\footnote{in
  this case $G_{\mu\nu} = \kappa\,T_{\mu\nu}$ lead to $0 = 0$ without
  involving $G$}, and is necessarily linked to an empty world. But,
without anything there is no physical world and anything generates
spacetime.

Here, we also like to point out that Newton's gravitational constant
$G$ is not replaced by a different natural constant in Einstein's
General Relativity, i.e. the theory of relativistic gravity. Both
theories, Newton and GR, although conceptional completely different,
rest fundamentally on the gravitational constant, where $G$ is not a
linearised version of a GR-type constant (but Newton's force-type
formulation is). \corr{Using dimensional analysis,
  $[G] = $~Force$^{-1} $Velocity$^4 = F^{-1}\,V^4$, which is
  consistent with $[G] = M^{-1}\,L^3\,T^{-2}$.}  Based on the facts,
that $G$ is a fundamental non-linearised natural constant and solely a
combination of force and velocity, we argue that $G$ is implicitly a
relativistic quantity, as the 'relativity-indicator' $c$ is the only
natural quantity with dimension $V$ (and a Lorentz scalar). Then, the
'relativistic' notation would be $\kappa = 8\pi\,G/c^4$ as it appears
in Einstein's field equation~\footnote{Keep in mind that, Newton:
  $\vF_{\sm{grav}} = - G\,m_1 m_2\,\hat{\vr}/r^2$ or
  $\Delta\Phi = 4\pi\,G\,\rho$, and GR:
  $G_{\mu\nu} = 8\pi\,G/c^4\,T_{\mu\nu}$.}. The constant $\kappa$
explicitly includes $c$ and has the dimension of force. In turn, this
can be used to define a natural unit of force, i.e.
$f_\sm{u} \equiv c^4/G$. Furthermore, as GR is a covariant,
i.e. Lorentz invariant, formulation of gravity which manifests in
Einstein's field equation, it makes $G$ a Lorentz scalar as well. This
supports that $G$ is an universal natural constant, as natural
constants necessarily must be Lorentz scalars.




\subsection{Fine-structure constant}

The fine-structure constant, or Sommerfeld constant, is a big mystery
in physics. Many famous physicists thought about its origin and
couldn't make sense of it\footnote{e.g., \it{It has been a mystery
    ever since it was discovered more than fifty years ago, and all
    good theoretical physicists put this number up on their wall and
    worry about it.} \citep[in 'Loose Ends' of
  ][]{Feynman85}}. Unfortunately, we are not resolving this puzzle,
but look at it from another prospective.

From the above discussion using arguments based on dimensional
analysis, we could derive the structural form of $\afs$ if we link the
number of fundamental dimensions \corr{($L$, $T$ and $M$)} to the
number of the fundamental naturals constants \corr{$G$, $c$, $\hbar$
  and $e$} (see Eqs.~(\ref{eq:const_relation}) and
(\ref{eq:fine_structure})). This result is based on these very simple
assumptions independent on physical details. In turn, we can use this
outcome to justify our approach.

Otherwise, to eliminate one of the natural constants $e$ or $\hbar$ based
on Eq.~(\ref{eq:const_relation}), the combination $e^2/\hbar\,c$ should be
unity or a mathematical constant (like $\pi$ or $\sqrt{2}$). The
fine-structure constant $\afs$ has an underivable value and therefore
does not reduce the amount of information we need to redundantly
link the elementary charge $e$ and Planck's constant $\hbar$. So far,
there is no explanation for this number and no mathematical function
or numerological combination can be used to deduce $\afs$ \citep[see
e.g.][for past struggle on this issue]{Kragh2003b}

With measurements of $\afs$ one can substitue
the elementary charge as
\be
e \equiv \sqrt{\afs\,\hbar\,c} \approx 0.085 \, ,
\label{eq:e_by_alpha}
\ee
where the number follows if we use the units $\hbar = 1$ and $c =
1$. The relation of the elementary charge to Planck's constant via the
fine-structure constant manifests again that $e$ is a quantum-based
quantity (as $\hbar$-related quantities are) and a relativistic
Lorentz scalar. Hence, $\afs$ can be used to replace the elementary
charge $e$ in the list of natural constants. Although $\afs$ is
dimensionless, it does not reduce the number natural constants. So
far, $\afs$ is an unexplained and undeducible quantity.

\subsection{Use of other units for the fundamental dimensions}

As mentioned in the introduction, we have the freedom to choose
different 'units' for our three fundamental dimensions without
changing the results discussed here. For instance, choosing Energy
($E$), Velocity ($V$) and Length ($L$) instead of $M$, $L$ and $T$,
the dimensions of the natural constants become $[G] =
E^{-1}\,V^{4}\,L$, $[c] = V$, $[\hbar] = E\,V^{-1}\,L$ and $[e] =
E^{1/2}\,L^{1/2}$. Then, the adapted relation of Eq.~(\ref{eq:const_relation}) is
\be
[G]^{x_G} \,
[c]^{x_c} \,
[\hbar]^{x_h} \,
[e]^{x_e} =
  E^0\,V^0\,L^0 \, .\nonumber
\ee
Nevertheless, the solution of this equation in terms of the exponents
$x_G$, $x_c$, $x_h$ and $x_e$
is the same than shown earlier. Namely, we find $x_G = 0$ and
$e^2/\hbar c$ is just a number also for this $E$-$V$-$L$ unit system. This
result is not surprising, but it should be stressed again that our
conclusions discussed in this paper are independent on the choice of
the fundamental dimensions and are based only on the {\it number} of
those. 
 
Another question is whether dimensionless numbers can replace the
concept of fundamental dimensions and get rid of them. Dimensionless
numbers from counting events are used to define specific units, like
the Second for the SI unit system, which is defined to be 9192631770
ticks of the ground state hyperfine transition of the caesium-133
atom. \corr{Although dimensionless, this number does not imply that we
  can remove Time $T$ the from our list of fundamental dimensions.
We still need the information on how to measure time and define time
units.} It is {\it information} and measurability that gives rise for
the necessity of fundamental dimensions and not whether this
information is given in dimensionless units or other units.


The same applies for the frequently used 'natural' unit system where
$\hbar = 1 = c$ and, less frequent, $G = 1$. Although a very practical
unit system, in particular to convert e.g. length, time and mass to
energy, it does not imply that the number of fundamental dimensions
can be reduced. We still need the information of dimensions and
eventually specific units to quantify and measure physical
phenomena. Even simple experiments like a pendulum can't be quantified
with dimensionless numbers only. A pendulum is defined by (at least)
two independent information, the amplitude and the frequency. So, to
quantify a pendulum we need the information 'amplitude' and the
information 'frequency'. That those quantities can be measured by,
e.g. the dimensions 'Length' and 'Time' is not essential, but
essential is that we need two independent information. 

Again, this work is based on the fact that nature can be described
with only three independent fundamental dimensions, where the actual
definitions and the associated units are not decisive. 


\section{Redefinition of fundamental natural constants}

The above results also show, that the natural constants are
necessarily connected to fundamental dimensions. Dimensionless
constants can only be derived from the combination $e^2/\hbar c$. Within
this context the question arises how unique are the definitions of
natural constants and, can they be redefined, best in terms of
dimensionless numbers?


For instance, one can use another definition for the Gravitational
constant $G$, e.g. as discussed in \cite{Wutke23} introducing a
'Rationalized Metric System' (RMS), where $G$ is subsumed into a
gravitational mass $m_g \to \sqrt{G}\,m_I$ and $m_g$ being the
rationalized Newtonian mass and $m_I$ the inertial mass. Then,
Newton's gravitational force reads analog to the Coulomb force as
$\vF_{\sm{grav}} = - m_{g,1} m_{g,2}\,\hat{\vr}/r^2$ and
$m_I\,\va = \vF_{\sm{grav}}$ \citep[this is also the notation used by
Isaac Newton in his famous 'Principia', see also][and references
therein]{Haug22} without the coupling constant $G$. Nevertheless, the
equivalence principle, which is the foundation of General Relativity,
demands $m_I \propto m_g$ with a universal constant to fix the linear
relation of the inertial and gravitational mass, hence
$m_I = \gamma\,m_g$, $\gamma = $ const.  Obviously, the dimensions of
$m_g$ and $m_I$ are different in the redefined case with
$[m_g] = M^{1/2}\,L^{3/2}\,T^{-1}$ (if we sensibly assume that $m_I$
has the dimension of mass $M$) and $\gamma$ has the dimension of
$[\gamma] = M^{1/2}\,L^{-3/2}\,T$. To determine the value of $\gamma$
one can also use Cavendish-type torsion experiments. Not surprising,
the result will be $\gamma = G^{-1/2}$ and hence, non of our
conclusions will change using $m_g$ and $\gamma$ instead of $G$.

What are the advantages of those redefined quantities. First we don't
need a coupling constant to express the gravitational force and
achieve an explicit analogy to the Coulomb force. Furthermore, the
different dimensions of $m_I$ and $m_g$ shows explicitly the
conceptual difference of the inertial and gravitational mass in
Newton's equation of motion. However, General Relativity {\em
  requires} that there is no conceptual difference between those
masses but they {\em must} be equivalent. The consequence is, that
gravity is not a force but a geometrical phenomenon. Therefore, we
argue that the common definition of $G$ leads to the appropriate and
most simple description of the physical laws. Hence, $G$ is the most
'natural' constant in this game to date and therefore a natural
physical constant. Of course, we are free to use other definitions for
$G$, but any other way lead to more elaborate and less intuitive
descriptions of physics.

The same is true for the common use of the other natural constants,
$\hbar$ and $c$ ($e$ is linked to $\hbar$ and $c$ via the fine-structure
constant). As long as we are not able to deduce the natural constants
from first principles, for instance from a 'Grand Unified Theory'
(GUT), the way that they are implemented up to date is the most
natural and meaningful implementation (see also Eqs.(\ref{eq:G_dims})
to (\ref{eq:e_dims}) for the associated dimensions). Hence our results
are as general as physical laws are valid till today.

\section{On the number of fundamental natural constants}
\label{sec:phys_consts}



This work is based on the assumption that there are only {\em four}
natural constants. Those are the gravitational constant $G$, the speed
of light $c$, Planck's constant $\hbar$ and the charge of the electron
$e$~\footnote{Note again that $e$ can be replaced by the
  fine-structure constant $\afs$ (in combination with $\hbar$ and $c$, see
  Eq.~(\ref{eq:e_by_alpha})).}. Our findings, the form of the
fine-structure constant, Eq.~(\ref{eq:fine_structure}), and the
special role of gravity, Eq.~(\ref{eq:nograv}), show that we are on
the right track and our assumptions make sense. \corr{Otherwise,
  physics needs more than just those four natural constants. In what
  follows, we state arguments for our selection of constants.}

\corr{The natural constants $G$, $c$, $\hbar$ and $e$} are connected
to the different fundamental areas of physics: gravity, relativistic
physics, i.e. the propagation of massless particles, quantum mechanics
and electromagnetism. All physical descriptions of interactions fall
into those fundamental areas. So far physicists (or other creative
people) are not aware of any phenomenon of nature which are not part
of those fundamental physical areas \citep[see also the contribution
by Okun in][]{Duff02}.

\corr{For instance, the electroweak interaction or the strong force
  are subareas of quantum mechanics. For these, further parameters}
are necessary to express the associated physical laws (at least at the
present state of research, where we don't have a satisfying unified
theory which might also deduce those parameters and natural
constants). We need, for example, conserved quantum numbers and masses
of particles. In principle, one could declare masses, like the mass of
the electron, as fundamental natural constants. But in contrast to the
charge of an electrons, its mass is not a quantised universal
quantity, hence one needs many different 'fundamental' masses for all
the different particles of the standard model of elementary
particles. Therefore, particle masses are commonly considered
{\it parameters} rather than fundamental natural constants. Similarly are
other coupling constants, like Fermi's constant of the weak
interaction or the QCD coupling constant, considered to be parameters
and not fundamental natural constants. Those are either connected to
particle masses (e.g. the W-boson mass in the case of Fermi's
constant) or are determined by perturbation theory as 'running
constants' and might be linked
to 'unknown' physics~\footnote{Here we subsume also proposed theories,
  like supersymmetry, string or M-theory, in 'unknown' physics.}.





We also like to mention, that the Boltzmann constant $\kb$ is not a
{\em fundamental} natural constant but a conversion factor to convert
temperature to energy, and hence has the dimensions of
Energy$\times$Temperatur$^{-1}$. Temperature is a very intuitive and
convenient proxy for energy, specifically internal energy which is the
result of random motions of atomic particles. The concept of
temperature and the very important field of thermodynamics is
based on statistical descriptions of underlying fundamental physical
processes, in particular Quantum Mechanics, but it is not a
fundamental area of physics itself.
Therefore, there is also no fundamental natural constant to associated
thermodynamics.


\section{Using natural constants to define fundamental unit systems}
\label{sec:unit_system}

Here we want to discuss possible natural unit systems based on the
mentioned fundamental natural constants, $G$, $c$, $\hbar$ and
$e$. Possible solutions are the Stoney units \citep{Stoney1881} and
the Planck units \citep[which were already proposed by Max Planck in
1899,][]{Planck1899}. Again, we use argument of dimensional analysis to
determine the basic units for mass, length, time and derived units.

Starting with mass $M$ we solve
\be
\left[G\right]^{x_{G,M}}\, [c]^{x_{c,M}}\, [\hbar]^{x_{h,M}}\, [e]^{x_{e,M}} = M
\label{eq:mass_from_constants}
\ee
for the exponents $x_{G,M}$,$x_{c,M}$, $x_{h,M}$ and
$x_{e,M}$, which solution is
\be
M = \left[G\right]^{-1/2}\,[c]^{x_{h,M}}\,[\hbar]^{x_{h,M}}\,[e]^{1-2x_{h,M}} =
\frac{[e]}{\left[G\right]^{1/2}}\, \left[\frac{\hbar\,c}{e^2}\right]^{x_{h,M}} \quad ,
\label{eq:mass_by_consts}
\ee
i.e., $x_{G,M} = -1/2$, $x_{c,M} = x_{h,M}$ and
$x_{e,M} = 1 - 2x_{h,M}$ with an undetermined exponent
$x_{h,M}$. Again, the term in the square brackets is the
fine-structure constant and does not contribute to uniquely determine
the connection of natural units and natural constants. Only
$x_{G,M}=-1/2$ is a unique solution of
Eq.~(\ref{eq:mass_from_constants}), which again indicates the special
role of gravity.

The mentioned natural unit systems, Stoney and Planck, are given by
$x_{h,M} = 0$ and $x_{h,M} = 1/2$, respectively.
But generally, one can identify the unit of mass as
\be
m_u \equiv \frac{e}{\sqrt{G}}\,
\left(\frac{\hbar\,c}{e^2}\right)^{x_{h,M}} \quad .
\label{eq:mass_unit}
\ee

To find the natural unit system for fundamental dimensions length $L$
and time $T$ based on $G$, $c$, $\hbar$ and $e$ we solve:
\bea
\left[G\right]^{x_{G,L}}\,  [c]^{x_{c,L}}\, [\hbar]^{x_{h,L}}\, [e]^{x_{e,L}} & = & L 
\label{eq:length_by_consts} \\
\left[G\right]^{x_{G,T}}\, [c]^{x_{c,T}}\, [\hbar]^{x_{h,T}}\,  [e]^{x_{e,T}} & = & T
\label{eq:time_by_consts}
\eea
which solutions are
\bea
L & = & \frac{\left[G\right]^{1/2}\,[e]}{[c^2]}\,
\left[\frac{\hbar\,c}{e^2}\right]^{x_{h,L}} \\
T & = & \frac{\left[G\right]^{1/2}\,[e]}{[c^3]}\,
\left[\frac{\hbar\,c}{e^2}\right]^{x_{h,T}} 
\eea
again with undetermined exponents $x_{h,L}$ and $x_{h,T}$.  Hence, the
general natural units are:
\bea
l_u & = & \frac{\sqrt{G}\,e}{c^2}\,
\left(\frac{\hbar\,c}{e^2}\right)^{x_{h,L}} \\
t_u & = & \frac{\sqrt{G}\,e}{c^3}\,
\left(\frac{\hbar\,c}{e^2}\right)^{x_{h,T}}
\label{eq:length_time_unit}
\eea

Note, for the solutions of Eqs.~(\ref{eq:mass_by_consts}),
(\ref{eq:length_by_consts}) and (\ref{eq:time_by_consts}) only the
$x_G$'s, the exponents linked to the gravitational constant $G$ are
uniquely determined. The specific appearance of the other natural
constants in a natural unit system can not be fixed with the arguments
of dimensional analysis.
This fact is in line with the discussion whether one can reduce the
number of natural constants as discussed in
Sec.~\ref{sec:dims_consts}: Gravity is certainly special and stands
out as it uniquely determines a natural unit system and {\em can not}
be removed from its definition. Otherwise, the other natural constants
can be removed from a natural unit system by an appropriate choice of
the $x_h$-exponents, as we will discuss below.

\subsection{Stoney and Planck unit system}


As mentioned above the Stoney and Planck unit systems emanate from
$x_{h,(M,L,T)} = 0$ and $x_{h,(M,L,T)} = 1/2$,
respectively. Explicitly, they are given by
\begin{itemize}
  \item Stoney unit system, $x_h = 0$:
\bea
m_{\sm{St}} & \equiv & \frac{e}{\sqrt{G}} \quad \approx 1.86\times 10^{-6}
                       \,\g  \nonumber \\
l_{\sm{St}} & \equiv & \frac{\sqrt{G}\,e}{c^2} \quad \approx 1.38\times
10^{-34}\,\cm \nonumber \\
t_{\sm{St}} & \equiv & \frac{\sqrt{G}\,e}{c^3} \quad \approx 4.61\times 10^{-45}\,\sec
\label{eq:Stoney_units}
\eea
\item Planck unit system, $x_h = 1/2$:
\bea
m_{\sm{Pl}} & \equiv & \left(\frac{\hbar\,c}{G}\right)^{1/2} \quad \approx 2.18\times
      10^{-5} \,\g \nonumber \\
      l_{\sm{Pl}} & \equiv & \left(\frac{\hbar\,G}{c^3}\right)^{1/2} \quad \approx 1.62\times
      10^{-33}\,\cm \nonumber \\
      t_{\sm{Pl}} & \equiv & \left(\frac{\hbar\,G}{c^5}\right)^{1/2} \quad \approx 5.39\times
                                                                   10^{-44}\,\s
\label{eq:Planck_units}
\eea
\end{itemize}

Both systems differ only by a factor of $\sqrt{\afs} \approx
11.7$. Both solutions reflect two 'asymptotic' cases, where the Stoney
units are independent on the Planck constant $\hbar$, whereas the
Planck units are independent on the elementary charge $e$. It is not
surprising, that Stoney didn't include the Planck constant $\hbar$ in
his unit system as it was not known that charges are quantised and
Quantum Mechanics was not yet invented when he published his studies
on the electron and natural unit system around 1880s. The relation of
those two unit systems was already discussed by Percy Williams
Bridgman \citep[pages 100 ff]{Bridgman22} based on the work by Lewis
and Adams \citep{Lewis1914}, without linking it to the today familiar
fine-structure constant which was introduced by Sommerfeld only in
1916 \citep[see also][for the pre-Sommerfeld connections to
$\afs$]{Kragh2003a}.

It should also be mentioned that both solutions are minimal in the
sense that they use only one common $x_h$-exponent for all three
units for mass, length and time. 


With the knowledge that the elementary charge $e$ is a quantum-based
natural constant and the subsequent one-to-one relation of $e$ with
Planck's constant via the fine-structure constant and the speed of
light ($\afs = e^2/\hbar\,c$), it is evident that both unit systems
incorporate Quantum Mechanics and hence represent all fundamental
physical characteristics and phenomena (note that, $e$ is also a
'relativistic' quantity due to the relation with $\afs$, and it is
directly linked to radiation). Nevertheless, the notation of the
Planck unit system Eq.~(\ref{eq:Planck_units}) brings together those
fundamental physical characteristics, GR and QM, a bit more obviously,
as all units are based on $G$, $\hbar$ and $c$, whereas in the Stoney unit
system $m_{\sm{St}}$ does not include $c$. This might be one reason
why the Planck unit system prevails and the Stoney system is never
used in modern physics. \corr{We also like to point out, that the
  usage of these unit systems is normally motivated by physics (as
  they link the key physical areas, see discussion in
  Sec.~\ref{sec:phys_consts}). Nevertheless, our formal approach based
  on dimensional analysis can reproduce those known unit systems if we
  fix the undetermined exponents $x_h$.}

\subsection{Alternative unit systems?}

From Eqs.~(\ref{eq:mass_unit}) and (\ref{eq:length_time_unit}) we can
think on alternatives to the Stoney and Planck units, which are units
without $\hbar$ and $e$, respectively. For instance, we
could define a unit system without the speed of light $c$, i.e.
\bea
m_{\sm{no}\, c} & \equiv & \frac{e}{\sqrt{G}} \quad = \quad
m_{\sm{St}} \sim 1.9\times 10^{-6}\,\g \nonumber \\
l_{\sm{no}\, c} & \equiv & \frac{\sqrt{G}\,\hbar^2}{e^3} \sim 2.6\times
10^{-30}\,\cm \nonumber \\
t_{\sm{no}\, c} & \equiv & \frac{\sqrt{G}\,\hbar^3}{e^5} \sim 1.2\times
10^{-38}\,\sec .\nonumber
\eea
Such an unit system would implicitly assume that nature is
fundamentally relativistic without explicitly addressing it
(nevertheless, as mentioned earlier $e$ is a 'relativistic' as well as a
'quantum' natural constant). It comes on the cost of including both
quantum concepts, $\hbar$ and $e$, at the same time, and, maybe worse, is
based on different $x_h$'s for each unit, namely $x_{h,M} = 0$,
$x_{h,L} = 2$ and $x_{h,T} = 3$. This unit system is neither intuitive
nor particular useful.

Any other unit system based on the concept of
Eqs.~(\ref{eq:mass_unit}) and (\ref{eq:length_time_unit}) will not
give any more perception compared to the Stoney and Planck
system. Nevertheless, we like to stress again, that no unit system
based on the natural constants can be derived without $G$, the
gravitational constant. Otherwise, as Quantum Mechanics can be
expressed either by the elementary charge $e$ or Planck's constant
$\hbar$, one can not design a unit system without $e$ and $\hbar$ at the same
time, i.e. without QM.

In conclusion of this discussion on alternative unit systems, the most
intuitive and reasonable natural unit system is the Planck system,
Eqs.~(\ref{eq:Planck_units}), as it explicitly highlights the three
fundamental physical concepts, Gravity, Relativity and Quantum
Mechanics.

\section{Conclusions}

Dimensional Analysis is a powerful tool in physics. It is helpful, for
instance, to find general dependencies, to make order-of-magnitude
estimates and to relate physical quantities \citep[for a general
discussion on Dimensional Analysis and examples
see][]{Bridgman22}. E.g., the $v^3$ dependency of the power of air and
fluid flows can be inferred from it or, together with natural
constants, the time of an object attracted by gravitationally by a
mass $M$ results in $t_{\sm{ff}} \sim R^{3/2}/\sqrt{G\,M}$ (which is
also Kepler's third law) just from arguments by dimensional
analysis. Including Planck's constant, i.e. analysing quantum
mechanical systems, one finds the Compton wave length
$\lambda = h/m c$ from dimensional analysis when we want to associate
a typical wavelength to a particle with mass $m$ (which is the
characteristic wavelength of the Compton effect).


In this article, we used Dimensional Analysis to explicitly combine
the natural constants \corr{($G$, $c$, $\hbar$ and $e$)} and the
fundamental dimensions \corr{($L$, $T$, and $M$)} of the physical
world. This approach leads to the composition and hence
the 'structural form' of the fine-structure constant, i.e.
$\alpha \sim e^2/c\,\hbar$, without any further
assumption. \corr{Otherwise,} 
dimensional analysis can't be applied to determine precise numbers and
hence, it does not resolve the puzzle of the value of the
fine-structure constant. 

Furthermore, we found that $G$ can not be expressed by other natural
constants, which supports that gravity is a peculiar area of physics
and can not be easily combined with quantum mechanics. The peculiarity
of gravity comes from the fact that it is not a force-type interaction
but is a result of the properties of Spacetime.









\end{document}